\documentclass[conference]{IEEEtran}
\IEEEoverridecommandlockouts

\def\BibTeX{{\rm B\kern-.05em{\sc i\kern-.025em b}\kern-.08em
    T\kern-.1667em\lower.7ex\hbox{E}\kern-.125emX}}

\usepackage{url}
\usepackage{balance}
\usepackage[htt]{hyphenat}
\usepackage{multirow}
\usepackage[inline]{enumitem}
\usepackage{listings}
\usepackage{graphicx}
\usepackage{xcolor}
\usepackage{longtable}
\usepackage[acronym]{glossaries}

\begin{document}

\title{\emph{Mascara}: \\A Novel Attack Leveraging Android Virtualization\\}

\author{\IEEEauthorblockN{Marco Alecci}
\IEEEauthorblockA{\textit{Department of Information Engineering} \\
\textit{University of Padua}\\
Padua, Italy \\
 marcoalecci98@gmail.com}
\and
\IEEEauthorblockN{Riccardo Cestaro}
\IEEEauthorblockA{\textit{Department of Mathematics} \\
\textit{University of Padua}\\
Padua, Italy \\
riccardo.cestaro.1@studenti.unipd.it}
\and
\IEEEauthorblockN{Mauro Conti}
\IEEEauthorblockA{\textit{Department of Mathematics} \\
\textit{University of Padua}\\
Padua, Italy \\
conti@math.unipd.it}
\and
\IEEEauthorblockN{Ketan Kanishka}
\IEEEauthorblockA{\textit{Department of Computer Science and Engineering} \\
\textit{Indian Institute of Technology (BHU)}\\
Varanasi, India \\
ketan.kanishka.cse15@iitbhu.ac.in}
\and
\IEEEauthorblockN{Eleonora Losiouk}
\IEEEauthorblockA{\textit{Department of Mathematics} \\
\textit{University of Padua}\\
Padua, Italy \\
eleonora.losiouk@unipd.it}
}

\maketitle

\begin{abstract}
Android virtualization enables an app to create a virtual environment, in which other apps can run. 
Originally designed to overcome the limitations of mobile apps dimensions, malicious developers soon started exploiting this technique to design novel attacks. As a consequence, researchers proposed new defence mechanisms that enable apps to detect whether they are running in a virtual environment. 

In this paper, we propose \emph{Mascara}, the first attack that exploits the virtualization technique in a new way, achieving the full feasibility against any Android app and proving the ineffectiveness of existing countermeasures. \emph{Mascara} is executed by a malicious app, that looks like the add-on of the victim app. As for any other add-on, our malicious one can be installed as a standard Android app, but, after the installation, it launches \emph{Mascara} against the victim app. The malicious add-on is generated by \emph{Mascarer}, the framework we designed and developed to automate the whole process. Concerning \emph{Mascara}, we evaluated its effectiveness against three popular apps (i.e., Telegram, Amazon Music and Alamo) and its capability to bypass existing mechanisms for virtual environments detection. We analyzed the efficiency of our attack by measuring the overhead introduced at runtime by the virtualization technique and the compilation time required by \emph{Mascarer} to generate 100 malicious add-ons (i.e., less than 10 sec). Finally, we designed a robust approach that detects virtual environments by inspecting the fields values of \textit{ArtMethod} data structures in the Android Runtime (ART) environment. 
\end{abstract}

\begin{IEEEkeywords}
Mobile Security; Android; Virtualization Technique
\end{IEEEkeywords}

\section{Introduction}
\label{sec:introduction}

Google Play Store is a broad and wild environment, presently counting more than three millions Android apps~\cite{AppBrain}, published by companies, academies or passionate developers. The openness of the Android platform encourages many people to contribute to the growth of the Play Store. Unfortunately, some of those people exploit it as an attack vector to disseminate malwares. Google has adopted different defence mechanisms (e.g., Google Bouncer~\cite{Bouncer} introduced in 2012; Google Play Protect~\cite{PlayProtect} released in 2017) to verify the apps immediately after their publication or update. Despite those efforts, malware developers still manage to overcome such detection mechanisms, as it is confirmed by the 870,617 malicious installation packages found in 2019 according to this Kaspersky Lab analysis~\cite{KasperskyMalwareNumber}. 

Among the different approaches used to design a malware, attackers have recently started exploiting the virtualization technique. When used in the Android Operating System (OS), this technique gives the opportunity to an app (i.e., \emph{Container App}) to define a new environment, separate from the Android default one, that is able to run any app (i.e., \emph{Plugin App}), while fully preserving its functionalities. The virtualization technique relies on the dynamic code loading and on the dynamic hooking. The dynamic code loading enables an Android app to bypass the 64K reference limit~\cite{multiDex} by loading the external code contained in a Dalvik EXecutable (DEX) file, a Java ARchive (JAR) file or even in the Android Package (APK) of another app (the Android app bytecode can reference up to 65,536 methods, among which the Android framework methods, the library methods and the app own methods). The dynamic hooking allows an Android app to intercept at runtime any call to the Android framework. This approach is not officially supported by Google, but several solutions are already available~\cite{ARTDroid, PluginBook, HookingTechniques}. 

The popularity of the virtualization technique, confirmed by the number of downloads of the apps on the Play Store implementing it~\cite{ParallelSpace,parallelAccounts,multipleAccounts}, is given by its main use case scenario, i.e., running  multiple instances of the same app on a single device: if a user has two separate Telegram accounts and wants to use them simultaneously, he can have the first running in the Android environment and the second in one of the virtualized contexts. The same goal might be achieved by creating different users on the same device through the \emph{Multi-user system behavior}, introduced in Android 5.0~\cite{multipleUsers}. However, this approach has a significant usability limitation: while using a specific user account, the notifications of apps running under a different user are not displayed. 

Previous works~\cite{PluginCatastrophe, DroidPill, AppInTheMiddle, ParallelSpaceTravelling, DiPrint} have already identified attacks exploiting the virtualization technique and even very popular apps have been already the target of new virtualization-based malwares~\cite{TwitterMalware, WhatsAppMalware}. Thus, researchers~\cite{DoNotJail, DiPrint, ParallelSpaceTravelling, AppInTheMiddle, Antiplugin} have soon started designing novel detection mechanisms, specifically customized to detect virtual environments, which are different from the more well-known emulated environments. However, the attacks identified so far rely only on built-in features of the virtualization technique, provided by all frameworks supporting the generation of Android virtual environments~\cite{DroidPlugin, VirtualApp, DynamicAPK}. On the contrary, \emph{Mascara} is the first attack that relies on a customization process applied to the original virtualization framework and repeated everytime for each single victim app. Such customization process not only enhances the probability to lure users to install the app that will launch \emph{Mascara}, but it also allows to bypass the existing detection mechanisms for virtual environments. Moreover, \emph{Mascara} is the first attack that can be launched against any Android app and the framework we developed, \emph{Mascarer}, automatically generates the malicious add-on of a victim app. Thus, \emph{Mascara} highlights new features of the virtualization technique that can be exploited to generate undiscovered attacks, even more threatening than the existing ones. \emph{Mascara} is launched by a malicious app, that is customized to look like the legitimate add-on of a popular app, while it is responsible for the generation of a virtual environment. Its purpose is to extract sensitive data from the victim smartphone by leveraging on the generation of a virtual environment, where the victim app and a malicious app run simultaneously.

\textbf{Contribution.} 
\begin{itemize}
    \item We designed and developed \emph{Mascara}, a novel virtualization-based attack that bypasses existing defence mechanisms, and \emph{Mascarer}, a framework for the automatic generation of the malicious add-on launching \emph{Mascara};
    \item We evaluated the effectiveness of \emph{Mascara} against Alamo, Amazon Music and Telegram and the runtime overhead introduced by the virtual environment. We also evaluated the efficiency of \emph{Mascarer} by measuring the compilation time required to generate the malicious add-ons of 100 popular apps;
    \item We executed the existing defence mechanisms (i.e., \emph{AntiPlugin}~\cite{Antiplugin} and \emph{DiPrint}~\cite{DiPrint}) in \emph{Mascara} and proved that our attack is resilient to such approaches;
    \item We finally developed a robust detection mechanism to detect Android virtual environments by inspecting the fields values of \textit{ArtMethod} data structures in the Android Runtime (ART) environment.
\end{itemize}

\textbf{Organization.} The paper is organized as follows: in Section~\ref{sec:background}, we describe the virtualization technique applied to the Android OS, as well as the features of the DroidPlugin framework used to design \emph{Mascara}; in Section~\ref{sec:related_work}, we introduce the related work, while in Section~\ref{sec:persona_attack}, we illustrate the \emph{Mascara} attack. In Section~\ref{sec:design} and Section~\ref{sec:implementation}, we describe the design and implementation of the \emph{Mascarer} framework; in Section~\ref{sec:evaluation}, we evaluate the efficiency of \emph{Mascarer} and the effectiveness of \emph{Mascara}. In Section~\ref{sec:attack_detection}, we prove that \emph{Mascara} is resilient to the existing defence mechanisms. Finally, in Section~\ref{sec:discussion}, we discuss about limitations and illustrate our new detection mechanism, before concluding the paper.

\section{Background on Android Virtualization} 
\label{sec:background}
Android is a multi-process, Linux-based OS, that executes apps in isolated processes, giving each one a unique Linux UID, its own address space and a private directory. By enabling apps to expose a virtual environment in which other apps can run, the virtualization technique~\cite{Antiplugin,DroidPill} breaks the current Android security model. DroidPlugin~\cite{DroidPlugin}, VirtualApp~\cite{VirtualApp} and DynamicAPK~\cite{DynamicAPK} are the most well-known frameworks supporting the virtualization technique and they differ according to whether their virtual environment can: (i) run one or multiple apps simultaneously; (ii) run apps not installed on the device; (iii) run any Android app or only a specific set of apps.  

\begin{figure}[h]
	\centering
	\includegraphics[width=0.45\textwidth]{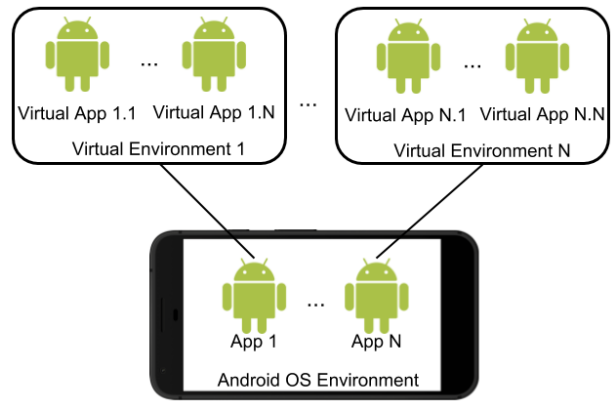}
	\caption{Multiple virtual environments created through DroidPlugin.}
	\label{fig:AndroidVirtualization}
\end{figure}

To design and develop \emph{Mascara}, we evaluated all above-mentioned frameworks and we chose to use DroidPlugin, since it allows to build the most damaging attack in comparison to the other frameworks: its virtual environment can execute any Android app, even multiple ones at the same time, including those not installed on the device, and it can be installed as a generic Android app, without requiring additional privileges enabled on the device (e.g., root privileges). In Fig.~\ref{fig:AndroidVirtualization}, we show a smartphone with two Android apps using DroidPlugin to generate each one its own virtual environment, in which other apps can run. 

\begin{figure}[h]
	\centering
	\includegraphics[width=0.45\textwidth]{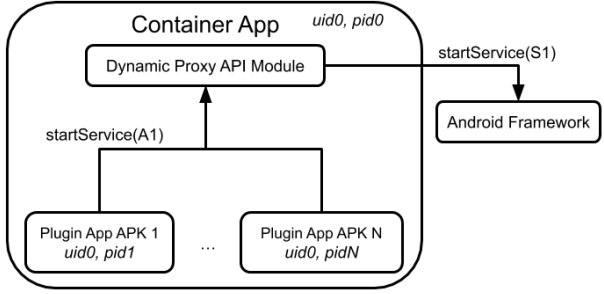}
	\caption{DroidPlugin internal architecture.}
	\label{fig:droidplugin_architecture}
\end{figure}

Fig.~\ref{fig:droidplugin_architecture} shows the internal architecture of an app generating a virtual environment through the DroidPlugin framework. The architecture involves the following components: a \emph{Container App}, several \emph{Plugin Apps} and a \emph{Dynamic Proxy API Module}. The \emph{Container App} is the main component, since it is responsible for the virtual environment generation through the dynamic code loading and the dynamic hooking. The \emph{Plugin Apps} are the apps running in the virtual environment, while the \emph{Dynamic Proxy API Module} contains the logic to implement the dynamic hooking through the Java Dynamic Proxy~\cite{DynamicProxy}. 

Once the \emph{Container App} has created the virtual environment, it forks its process (i.e., \emph{pid0}) into as many processes as the \emph{Plugin Apps} that it is going to run. The \emph{Container App} and the \emph{Plugin Apps} will share the same UID (i.e., \emph{uid0}), but each one will run in its own process. Then, the \emph{Container App} loads the executable files of the \emph{Plugin Apps}, saved in their APK files, and execute them in their dedicated processes. During the runtime execution, the \emph{Container App} has to manage the lifecycle of \emph{Plugin Apps} components. Android apps contain several components (i.e., \texttt{Activities}, \texttt{Services}, \texttt{Broadcast Receivers} and \texttt{Content Providers}), which need to be declared in the \texttt{AndroidManifest.xml} file so that the Android OS can register them at the installation time. The management of each component involves an interaction between the Android app and the Android OS: for example, to show an \texttt{Activity} on the screen of the smartphone, an app has to send a request to open that \texttt{Activity} to the Android OS and it has to receive a reply containing the details of the device where the app is running (e.g., the dimension of the screen). When the \emph{Plugin Apps} run in a virtual environment, the Android OS receives all requests sent by them as if they were sent by the \emph{Container App}. Thus, the \emph{Container App} is expected to declare the same components as the \emph{Plugin Apps} running in its virtual environment. Thanks to the \emph{Dynamic Proxy API Module}, DroidPlugin is able to execute any Android app without customizing each time the \emph{Container App}. As a matter of fact, the \emph{Container App} first declares a generic number of stub components and, then, it can rely on the \emph{Dynamic Proxy API Module} to intercept each request/reply going towards or coming from the Android OS to dynamically change the name of the component. This approach is used for \texttt{Activities}, \texttt{Services} and \texttt{Content Providers}, while \texttt{Broadcast Receivers} are parsed from the \emph{Plugin App} \texttt{AndroidManifest.xml} file and dynamically registered at runtime. The same approach is also followed for the permissions: the \emph{Container App} declares all existing Android permissions.

\section{Related Work}
\label{sec:related_work}
Previous works address the Android virtualization technique from three different points and can be divided in three groups, accordingly: the ones focusing on the exploitation of the virtualization technique to develop new attacks; the ones proposing novel detection mechanisms for virtual environments; the ones proposing the virtualization technique as a novel defence mechanism. 

\textbf{Attacks exploiting Android virtualization.} In a virtual environment, the attacks that can be performed are usually executed by the \emph{Container App} or by the \emph{Plugin App}~\cite{PluginCatastrophe, AppInTheMiddle, DiPrint, ParallelSpaceTravelling}. Concerning the former, previous works identified the following exploitations: 
\begin{itemize}
    \item The \emph{Container App} creates a virtual environment, where both benign and malicious apps are executed;
    \item The execution of an app running in the native Android OS is stopped and resumed within the virtual environment under the control of the \emph{Container App};
    \item The \emph{Container App} might download malicious code and dynamically execute it at runtime;
    \item The \emph{Container App} applies new hooks at runtime on the \emph{Plugin App} to make it execute malicious code (i.e., the new \emph{repacking attack}).
\end{itemize}

Concerning the \emph{Plugin App}, the possible attacks are as follows: 
\begin{itemize}
    \item \emph{Plugin Apps} can access files of each other;
    \item A \emph{Plugin App} shares the UID with the \emph{Container App}, which means also the same permissions;
    \item A malicious \emph{Plugin App} can tamper with the executable files or encrypt files of another  \emph{Plugin App};
    \item A \emph{Plugin App} can launch a phishing attack to steal user input provided to another  \emph{Plugin App} running in foreground.
\end{itemize}

In addition to the above-mentioned attacks, two popular apps were already found to be the target of a virtualization-based attack. In the first one~\cite{TwitterMalware}, a dual-instance app enables users to run a second instance of Twitter. Besides creating a virtual environment through \emph{VirtualApp}, the malware hijacks user inputs. The second malware aims at distributing a malicious, updated version of the WhatsApp app~\cite{WhatsAppMalware}. This is definitely similar to the original one, besides the additional malicious code crafted to steal users' sensitive data. After bypassing the Play Store checks, the developer managed to release his fake version with the same publisher name as the original WhatsApp (i.e., WhatsApp Inc.) and to have it downloaded more than one million times before its removal from the Store. 

All the attacks previously illustrated exploit the built-in features provided by the existing frameworks that support the generation of Android virtual environments. Thus, such attacks are not robust enough and can be detecting by anti-virtualization approaches.

\textbf{Android virtualization detection mechanisms.} The defence mechanisms proposed so far~\cite{Antiplugin, DiPrint, ParallelSpaceTravelling, AppInTheMiddle, DoNotJail} detect virtual environments by inspecting the following features: the permissions declared by the \emph{Container App} and shared with \emph{Plugin Apps}; the number and names of the components declared by the \emph{Container App}; the app process name; the information about the internal storage; the data sharing among \emph{Plugin Apps} and the stack trace of the \emph{Plugin App}. All such approaches are not effective, since a malicious \emph{Container App} can easily introduce new hooks on the Android and Java API, as well as on the native code, to modify their return values.

\textbf{Other usages of Android virtualization.}
NJAS~\cite{NJAS} and Boxify~\cite{Boxify} provide two interesting defence mechanisms against Android malwares. The first one builds a virtual environment where malwares run, thus preventing them from gaining control of the user phone. Boxify~\cite{Boxify} builds sandboxed environments, where malicious apps run in isolated processes, thus, preventing apps from requiring permissions and accessing the associated resources. Other previous works used virtualization to monitor the apps runtime behaviour~\cite{Afonso2016GoingNU, Davis12, ZhouHybrid}, to enhance the security level of virtual machine~\cite{Aurasium} or to support the hot patch feature~\cite{InstaGuard}. 

\section{\emph{Mascara} Attack}
\label{sec:persona_attack}
In this section, we provide an overview of \emph{Mascara} (Section~\ref{sec:mascara_overview}) and a detailed description of the internal architecture of the malicious add-on (Section~\ref{sec:malicious_add_on_architecture}). 

\subsection{Overview of \emph{Mascara}}
\label{sec:mascara_overview}
\emph{Mascara} refers to the Venetian-Italian word, which means mask and recalls the main purpose of our malware: stealing sensitive data from a smartphone by luring the user to install a malicious app, which looks like a legitimate one. \emph{Mascara} is launched by a malicious app, which is customized to look like the add-on of a popular app. Our choice to use the add-on technology as a vector for \emph{Mascara} is motivated by its numerous advantages. Generally speaking, add-ons are accessory components aimed at providing new functionality to existing systems. They have their own, separated logic, but they also strongly depend on the existing systems, without which they can not even work. In the Android environment, the add-on technology follows the same approach: there is a base app, which is itself a standalone app, and a variable number of add-ons, each one providing new features to the base app. Add-ons are Android apps, that have to run in the same mobile device, where the base app is installed, and that can be developed by different Android programmers with respect to the base app ones. An illustrative example of the add-on technology is provided by the ``Locus Map Free" app~\cite{Locus}, available on the Google Play Store. The base app is a navigation app that supports the offline usage of maps, but users can also download add-ons supporting satellite images~\cite{LocusAddon1}, augmented reality~\cite{LocusAddon2} or geo-caching~\cite{LocusAddon3}.
    
In the \emph{Mascara} threat model, the user just needs to install a single malicious app from the Play Store to become a victim. In fact, the attacker first uses our framework \emph{Mascarer} to customize the \emph{Container App} and generate the malicious add-on $V^\prime$ for the specific victim app $V$. Then, the attacker publishes $V^\prime$ on the Play Store as an important update for $V$ and the user, deceived by $V^\prime$ features, installs it on his smartphone. Since users typically do not investigate the apps they download~\cite{fakeApps}, this is a very simple requirement.
 
\begin{figure}[ht!]
	\centering
	\includegraphics[width=0.45\textwidth]{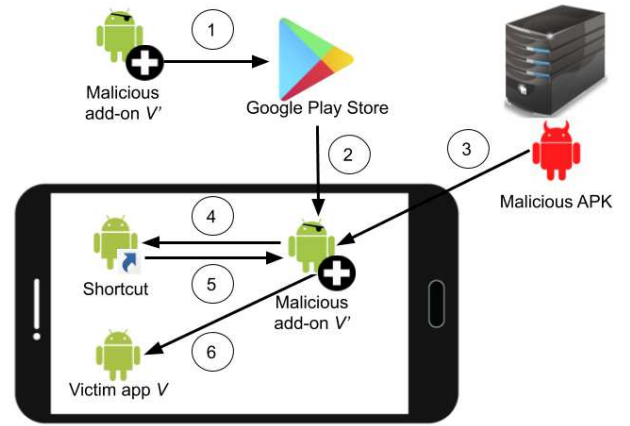}
	\caption{Complete workflow of \emph{Mascara}.}
	\label{fig:PersonaAttackScenario}
\end{figure}

In Fig.~\ref{fig:PersonaAttackScenario}, we illustrate the whole \emph{Mascara} attack. The first step is publishing the malicious add-on $V^\prime$, generated through \emph{Mascarer}, on the Google Play Store (step 1). Then, any Android user, already having the victim app $V$ installed on his device, can become a victim by installing $V^\prime$ (step 2). During the activation of $V^\prime$, this downloads a \emph{malicious APK} from a remote server and executes it in its virtual environment (step 3). At the same time, it creates a new shortcut in the Home screen of the device, that looks like a shortcut of $V$, while it executes $V^\prime$ (step 4). From now on, whenever the user presses that shortcut, $V^\prime$ is executed (step 5) to load $V$ (step 6) and to launch \emph{Mascara} against the user. 

To implement \emph{Mascara}, we identified the following design goals: (i) the malicious add-on performs the attack against a specific app; (ii) the malicious add-on builds a virtual environment, in which both the victim app and the malicious APK run; (iii) the malicious add-on requires no modification to the victim app source code; (iv) the malicious add-on requires no modification to the underlying Android OS; (v) the malicious add-on is a standard Android app, which can be installed by any user.
\subsection{Architecture of the Malicious Add-On}
\label{sec:malicious_add_on_architecture}

\begin{figure}[ht!]
	\centering
	\includegraphics[width=0.45\textwidth]{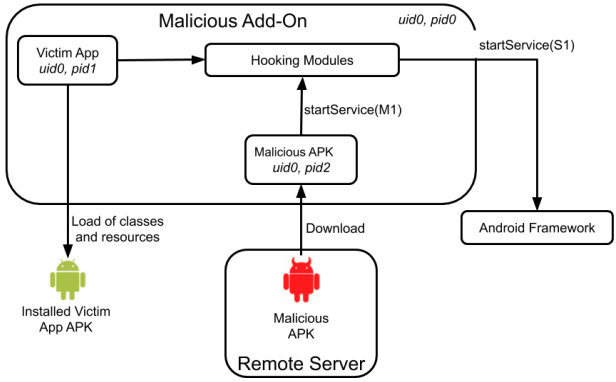}
	\caption{Architecture of the malicious add-on launching \emph{Mascara}.}
	\label{fig:malicious_add_on_architecture}
\end{figure}

The architecture of the malicious add-on, which is depicted in Fig.~\ref{fig:malicious_add_on_architecture}, encompasses the following components: the \emph{Victim App}, the \emph{Malicious APK} and the \emph{Hooking Modules}.  

The \emph{Victim App} component refers to the set of classes and resources belonging to the victim app, which is already installed on the device. As a matter of fact, the malicious add-on loads such files directly from the victim app APK, which is world-readable, and executes them in its virtual environment. Together with the \emph{Victim App}, the malicious add-on runs in its virtual environment also the \emph{Malicious APK}, which exploits the UID shared with the victim app to harm the final user in several ways (e.g., extracting or modifying sensitive data). At the installation time, the malicious add-on does not contain the \emph{Malicious APK}, which is instead downloaded at runtime from a remote server. Finally, the \emph{Hooking modules} include both the \emph{Dynamic Proxy API Module}, a DroidPlugin built-in module supporting the components lifecycle, and \emph{Whale}~\cite{whale}, an external hooking library. Both the \emph{Dynamic Proxy API Module} and \emph{Whale} are used to install new hooks and bypass existing defence mechanisms~\cite{Antiplugin, DiPrint}.

\section{\emph{Mascarer} framework design}
\label{sec:design}

In this section, we illustrate how \emph{Mascarer} has been designed to generate a malicious add-on for a given victim app. In particular, \emph{Mascarer} executes the following two procedures: (i) the customization process (Section~\ref{customization_process}); (ii) the exploitation of the virtualization technique (Section~\ref{malicious_exploitation_virtualization_technique}). 
\subsection{Customization Process}
\label{customization_process}

By default, the malicious add-on inherits the \texttt{AndroidManifest.xml} file from DroidPlugin, which declares all existing Android permissions and a set of predefined stub components, having the same repetitive structure. The purpose of the customization process is to make the malicious add-on resilient to the existing defence mechanisms by modifying its: (i) permissions, (ii) components and (iii) resources. More details about the defence mechanisms will be illustrated in Section~\ref{sec:attack_detection}.

\textbf{Permissions.} In addition to the defence mechanisms against virtualization, Android antiviruses~\cite{Puma} could also detect a malicious add-on declaring all Android permissions, since it violates the Android least-privilege security model (i.e., Android apps should declare the minimum amount of permissions required to work properly). Thus, during the customization process, \emph{Mascarer} adjusts both the malicious add-on and the \emph{Malicious APK} \texttt{AndroidManifest.xml} files to declare the same permissions as the victim app. This way, the malicious add-on will be able to run the victim app in its virtual environment and the \emph{Malicious APK} will be able to exploit the same permissions, without declaring additional ones. 

\textbf{Components.} As for the permissions, the malicious add-on first inherits all the stub components declared in the DroidPlugin \texttt{AndroidManifest.xml} file. Those stub components are exclusively used to change the names of the victim app's components at runtime. Thus, they have a very repetitive structure and meaningless names, which help defence mechanisms to detect frameworks supporting the generation of virtual environments. However, \emph{Mascarer} modifies both the malicious add-on and the \emph{Malicious APK} \texttt{AndroidManifest.xml} files. Concerning the malicious add-on, \emph{Mascarer} copies all the components of the victim app, preserving their original names, and modifies the additional components by giving them a name correlated with the victim app's name. With the same approach, \emph{Mascarer} customizes also the \emph{Malicious APK} components. 

\textbf{Resources.} Since the malicious add-on needs a custom launcher icon, when it is installed on the device, and the victim app launcher icon, to create the shortcut in user Home screen, \emph{Mascarer} copies those resources into the malicious add-on's ones. 
\subsection{Exploitation of the Virtualization Technique}
\label{malicious_exploitation_virtualization_technique}

\emph{Mascara} relies on two features of the virtualization technique: (i) the UID shared among all apps running within the same virtual environment; (ii) the hooking mechanism that allows to intercept APIs and change their return values at runtime.

\textbf{UID sharing among plugin apps.} In Android, sharing the same UID means also sharing the same permissions and having access to the same private directory. To exploit the shared UID, we rely on the DroidPlugin feature to run multiple plugin apps in the same virtual environment. Thus, our design involves that the malicious add-on runs only the victim app and a single \emph{Malicious APK} within the same virtual environment. Even if multiple apps can run simultaneously in the same virtual environment, only one can run in foreground and the others in background. According to this, we designed the \emph{Malicious APK} as an app with no graphical components (i.e., \texttt{Activities}), but only \texttt{Services}. In particular, by default, the \emph{Malicious APK} declares a set of \texttt{Services}, each one aimed at stealing/injecting specific data into the user smartphone. Some of such \texttt{Services} might need access to protected resources, thus, the \emph{Malicious APK} declares also the required permissions. During the customization process, \emph{Mascarer} modifies the \emph{Malicious APK} by removing all the \texttt{Services} besides the ones that can exploit the victim app's permissions. Once the \emph{Malicious APK} is downloaded from the remote server, the malicious add-on activates all its \texttt{Services}. Thanks to this design, at runtime the malicious add-on runs the victim app in foreground and all the malicious APK's \texttt{Services} in background. 

\textbf{Hooking mechanism.} By design, the \emph{Dynamic Proxy API Module} of DroidPlugin intercepts the outgoing requests and the incoming replies to dynamically change the names of the referenced components and guarantee the lifecycle of plugin apps' components. However, the same architecture can be used to install new hooks that either bypass existing defence mechanisms or damage the user. To install new hooks we both exploited the DroidPlugin hooking mechanism and the \emph{Whale} library. 
\section{\emph{Mascarer} framework implementation}
\label{sec:implementation}
In this section, we provide the technical details of the \emph{Mascarer} implementation concerning the customization process (Section~\ref{sec:customization_impl}), the first execution of the malicious add-on (Section~\ref{sec:starting_mal_add_on}) and the exploitation of the virtualization technique (Section~\ref{sec:virtual_env_mal_exploitation}). 

\subsection{Customization Process}
\label{sec:customization_impl}
We developed the customization process as a script, which performs the following steps: (i) updating permissions and features of the malicious add-on; (ii) customizing the malicious APK, which will be downloaded at runtime; (iii) updating the components of the malicious add-on; (iv) retrieving the necessary resources from the victim app to be copied into the malicious add-on. 

Considering the first step, the customization process relies on the Androguard Python tool~\cite{Androguard} to retrieve all permissions and features declared by the victim app and copy them into the malicious add-on \emph{AndroidManifest.xml} file. As a result, the add-on has the same permissions as the victim app plus the \texttt{INSTALL\_SHORTCUT} and the \texttt{KILL\_BACKGROUND\_PROCESSES} permissions used to create the fake shortcut in the Home screen and to stop the process of the victim app, respectively. 

During the second step, the customization process modifies the malicious APK so that it exploits only the permissions of the victim app. Thus, the permissions and associated \texttt{Services} of the malicious APK, which are not declared by the victim app, are removed. 

The purpose of the third step is to retrieve all components names of the victim app, as well as of the malicious APK, and create the equivalent components (i.e., names and types) in the \texttt{AndroidManifest.xml} file of the malicious add-on. In addition, since each component declared in the \texttt{AndroidManifest.xml} file needs a match with a .java file, the script creates the necessary ones in the add-on. Once again, to inspect the victim app and the malicious APK components, the customization process relies on the Androguard tool. Finally, the script customizes also the names of the remaining add-on components, which belong to the DroidPlugin framework and originally come with their default names (e.g., if Telegram is the victim app, the \texttt{PluginServiceManager} becomes \texttt{TelegramServiceManager}). 

During the fourth and last step, the script retrieves first the victim app name through Androguard and then its launcher icon from its resources. To achieve the second aim, the script gets first the name of the launcher icon through Androguard, then it unzips the victim app APK file and gets the icon under the predefined path. Since drawable images are not involved in the compilation process of an Android app, they do not need to be decompiled. 

\begin{figure}[h]
	\centering
	\includegraphics[width=0.45\textwidth]{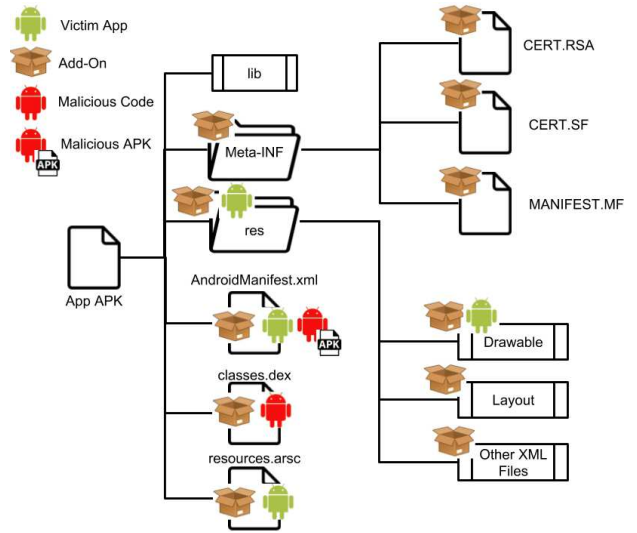}
	\caption{Malicious add-on internal components after the customization process.}
	\label{fig:customization}
\end{figure}

As shown in Fig.~\ref{fig:customization}, at the end of the customization process, the add-on encompasses components belonging to the victim app and to the malicious APK, as well as additional malicious code placed in its own logic. 

The customization process updates the following components inside the malicious add-on: the \texttt{res/} folder, the \texttt{AndroidManifest.xml} file, the \texttt{classes.dex} file and the \texttt{resources.arsc} file.  

The \texttt{res/} folder initially contains only the add-on uncompiled resources, among which the launcher icon and the XML layout file of the \texttt{Activity} shown to the user to confirm the add-on activation. In the same folder, the customization process saves the victim app launcher icon used to create the fake shortcut in the Home screen. 

After the customization, the \texttt{AndroidManifest.xml} file stores items belonging to the add-on, to the victim app and to the malicious APK as follows: items of the add-on (i.e., the \texttt{INSTALL\_SHORTCUT} and \texttt{KILL\_BACKGROUND\_PROCESSES} permissions, the \texttt{Activity} shown to the user during the first execution, all components belonging to the DroidPlugin framework and required to build a virtual environment); items of the victim app (i.e., all its permissions, components and features); items of the malicious APK (i.e., all its components). 

The \texttt{classes.dex} file contains the whole compiled code of an Android app. Considering the add-on, this file encompasses all classes belonging to the DroidPlugin framework plus the malicious hooks of the \emph{Dynamic Proxy API Module} and \emph{Whale} library to bypass the existing defence mechanisms. 

Finally, the \texttt{resources.arsc} file contains all information about resources, XML nodes and their attributes. After the customization, it comprises the add-on resources and the victim app launcher icon. 

\subsection{Starting the Malicious Add-on}
\label{sec:starting_mal_add_on}
During its first execution, the add-on performs the following activities: (i) stopping the victim app process already running on the phone through the \texttt{KILL\_BACKGROUND\_PROCESSES} permission; (ii) exploiting the \texttt{INSTALL\_SHORTCUT} permission to create a fake shortcut with the same icon and label of the victim app, but making it able to launch the malicious add-on; (iii) downloading the malicious APK from a remote server; (iv) executing the malicious APK.

\subsection{Exploitation of the Virtualization Technique}
\label{sec:virtual_env_mal_exploitation}
Our design of \emph{Mascara} involves that, at runtime, the add-on downloads a malicious APK exploiting the most dangerous permissions declared by the victim app (i.e., those allowing to access to or to modify sensitive data available on the phone). According to our design, the add-on runs simultaneously the victim app and the malicious APK in the same virtual environment. However, to prevent the user from noticing the existence of the malicious APK, we designed it as an app declaring only \texttt{Services}, without any graphical components. Thus, each \texttt{Service} aims at exploiting a specific Android permission and the whole set of \texttt{Services} declared by the malicious APK is everytime customized according to the victim app. As a result, at runtime the add-on launches all malicious APK \texttt{Services} and run them in background, together with the victim app. In our implementation, we developed separate \texttt{Services} for the following permissions: the \texttt{READ\_CONTACTS} permission to acquire the contacts already saved on the device; the \texttt{READ\_SMS} permission to access the SMS already saved on the device; the \texttt{RECEIVE\_SMS} permission for intercepting any new incoming SMS; the \texttt{READ\_PHONE\_STATE} permission for accessing the state of the phone; the \texttt{READ\_CALL\_LOG} permission for reading the user's call logs; the \texttt{CAMERA} permission for randomly taking pictures through the user phone; the \texttt{RECORD\_AUDIO} permission to record audio in background and the \texttt{ACCESS\_FINE\_LOCATION} to track user position. Finally, all above-mentioned \texttt{Services} exploit the \texttt{INTERNET} permission to send the collected information to our remote server. 

To further exploit the virtualization technique, we introduced new hooks, through DroidPlugin and \emph{Whale}, by identifying specific Android APIs and modifying their return values to bypass the existing defence mechanisms. 

\begin{figure}[h]
	\centering
	\includegraphics[width=0.47\textwidth]{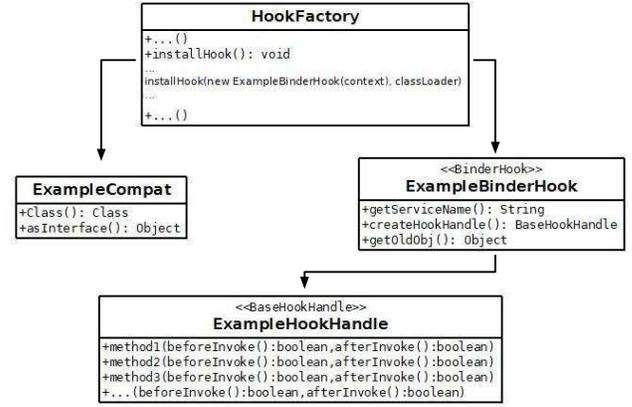}
	\caption{Classes involved in the DroidPlugin hooking process.}
	\label{fig:hooking}
\end{figure}

\textbf{DroidPlugin hooking mechanism.} If the hook is located in an Android system service, the hooking process requires the definition of three classes, which are also depicted in Figure~\ref{fig:hooking}:
\begin{itemize}
    \item Hook handle class (i.e., \emph{ExampleHookHandle}): it contains a set of inner classes, one for each method that needs to be hooked. Each inner class allows to specify the expected behaviour before and after the invocation of the method;
    \item Hook binder class  (i.e., \emph{ExampleBinderHook}): this class returns a proxy to the Android system service hooked, since the original one is cached, and an instance of the hook handle;
    \item Compatibility class  (i.e., \emph{ExampleCompat}): this class has to declare the static methods \texttt{Class()} and \texttt{asInterface()}. The name of the specified \texttt{Class} should be the same as the one of the \texttt{Interface} containing all the methods that need to be hooked. 
\end{itemize}
The hook can be installed through the method \texttt{installHook()} of the \texttt{HookFactory} class. 

If the hook is located in a different object other than an Android system service, the following scenarios might occur: 
\begin{itemize}
    \item If the \texttt{Class} of the object is an \texttt{Interface}, it is possible to use the \emph{Dynamic Proxy API} provided by Java.
    \item If the \texttt{Class} of the object is not an \texttt{Interface}, it is necessary to manually create a proxy class, by extending the targeted class.
    \item If the \texttt{Class} of the object is final, or if it has no public constructor, it is not possible to manually create a proxy class and another hooking location should be found.
\end{itemize}

\textbf{Whale hooking mechanism.} To install a hook in a different object other than an Android system service, we can rely on the \emph{Whale}~\cite{whale} library.
\emph{Whale} uses the \emph{Xposed}-style method hooking and it can bypass the Hidden API policy, the restriction applied to third-party apps not signed with the platform signature. Also, \emph{Whale} can modify the inheritance relationship between classes and the class an object belongs to. Through the \emph{Xposed}-style method hooking, it is possible to replace the entire \texttt{method} body or introduce new code before and after the original \texttt{method} invocation. When a \texttt{method} is invoked, the execution goes through the injected code and, then, to the original code. The method \texttt{findAndHookMethod} is used to find the target \texttt{method} and the callback can be:
\begin{itemize}
    \item \texttt{XC\_MethodHook}: the \texttt{class} that edits the fields before and after the original \texttt{method} invocation.
    \item \texttt{XC\_MethodReplacement}: the \texttt{class} that replaces the original \texttt{method} with a new one.
\end{itemize}

\section{\emph{Mascara} and \emph{Mascarer} evaluation}
\label{sec:evaluation}
In this section, we first analyze the efficiency of \emph{Mascarer} by measuring the compilation time required to generate the malicious add-ons for 100 popular apps. Then, we evaluate the effectiveness of \emph{Mascara} by implementing the attack against the Alamo, Amazon Music and Telegram apps and by measuring the overhead introduced at runtime by the virtualization technique. For all experiments, we used a Samsung Galaxy A5 (SM-A5000FU) running Android Marshmallow 6.0.1.

\begin{figure*}[ht!]
	\centering
	\includegraphics[width=1\textwidth]{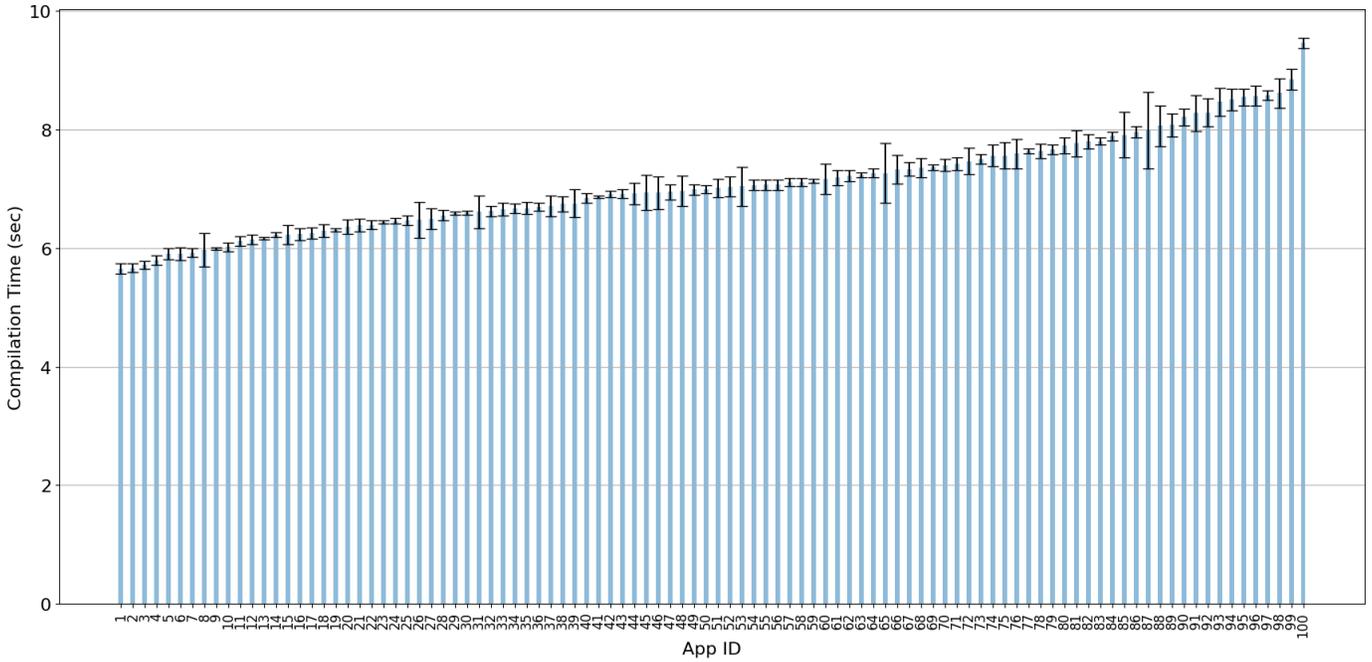}
	\caption{Average compilation time of 100 malicious add-ons.}
	\label{fig:average_time}
\end{figure*}

\textbf{Malicious Add-On Compilation Time.} Fig.~\ref{fig:average_time} shows the average compilation time (over ten rounds) required to generate the malicious add-ons for 100 popular apps available on the Google Play Store. To run the experiment, we used an ASUS VivoBook Pro N580GD-E4087T (Processor: Intel® Core™ i7 8750H; Memory: 16GB DDR4 2400 MHz SDRAM; Video card: NVIDIA® GeForce® GTX 1050 4GB DDR5; Storage: SSD M.2 512 GB SATA 3.0 + HDD 2.5” 1 TB 5400 RPM) running Linux Mint 19.3 Cinnamon. The complete list of the analyzed apps is available in Appendix~\ref{appendix:A}. The results show that for most of the malicious add-ons the compilation time is less than ten seconds. Considering the estimated amount of time that a smart attacker might need to manually generate the malicious add-on for a single app, this is a threatening result since \emph{Mascarer} allows to build the attack on a large scale set of apps with a limited amount of time.  

\begin{figure*}[ht!]
	\centering
	\includegraphics[width=0.5\textwidth]{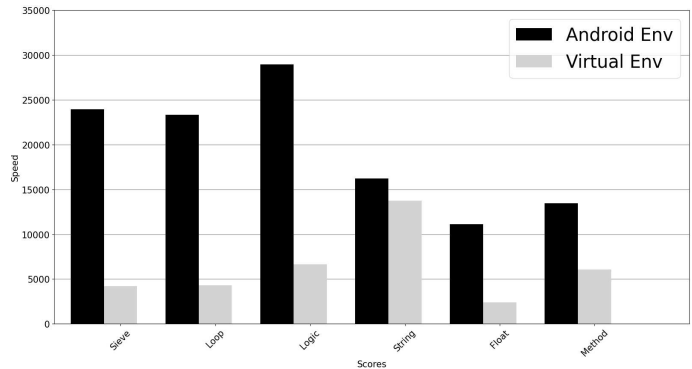}
	\caption{Performance comparison between the Android native environment and the Android virtualized environment.}
	\label{fig:CaffeineMark}
\end{figure*}

\textbf{Virtualization Technique Runtime Overhead.} To measure the runtime overhead introduced by the virtualization technique, we used the Java microbenchmark CaffeineMark (version 3.0)~\cite{CaffeineMark} ported on the Android platform. This benchmark runs a set of tests which allows a user to assess different aspects of virtual machine performance. The benchmark does not produce absolute values for the tests. Instead, it uses internal scoring measures, which are useful only in case of comparison with other systems. The overall score of CaffeineMark 3.0 is a geometric mean of all individual tests. To assess the virtual environment generated by DroidPlugin, we compared it with Stock Android system (version 6.0.1). The results are shown in Fig.~\ref{fig:CaffeineMark}. From the figure, we can observe that in the virtual environment the performances are reduced: the higher the score, the faster the system. 

\section{Defence Mechanisms against Android Virtualization}
\label{sec:attack_detection}
Among the existing defence mechanisms, we considered the \emph{AntiPlugin}~\cite{Antiplugin} and \emph{DiPrint}~\cite{DiPrint} solutions because their source code is available online and they provide a more comprehensive approach in detecting virtual environments with respect to other mechanisms~\cite{ParallelSpaceTravelling, AppInTheMiddle, DoNotJail}. In particular, we experimentally evaluated \emph{AntiPlugin} and \emph{DiPrint} by running both of them in \emph{Mascara} and verifying whether they can detect our virtual environment. In the rest of this section, we deeply describe each detection mechanism used by \emph{AntiPlugin} and \emph{DiPrint} and how we managed to bypass it.



\subsection{Feature for Virtual Environment Detection: Permissions}
By default, DroidPlugin declares all Android permissions to enable any app to run in its virtual environment. 

\textbf{Detection Mechanism 1.} Under VBA, the \emph{AntiPlugin App} is granted more permissions than expected, since it shares the same UID with the container app. Thus, the \emph{AntiPlugin App} can access to resources, which it should not be allowed to (e.g., the \texttt{BLUETOOTH} and \texttt{BLUETOOTH\_ADMIN} permissions are normal permissions, which are automatically granted at the installation and do not require a confirmation from the user). Since in \emph{Mascara} the malicious add-on declares the permissions of the victim app, the \emph{AntiPlugin App} is not able to access to different resources than the expected ones. 

\textbf{Detection Mechanism 2.} The \emph{AntiPlugin App} can check the permissions associated to the app with its own package name through the \texttt{PackageInfo} object~\cite{PackageInfo}. Under VBA, the \emph{AntiPlugin App}'s request triggers an error, since the app is not installed on the device. In \emph{Mascara}, the \emph{AntiPlugin App} is already installed and the request returns the expected list of permissions.  

\textbf{Detection Mechanism 3.} \emph{DiPrint} checks the undeclared permissions by calling the \texttt{checkCallingOrSelfPermission} API. Moreover, \emph{DiPrint} attempts to perform a set of actions that require dangerous permissions like \texttt{CAMERA,RECORD\_AUDIO and READ\_SMS}. In \emph{Mascara}, the malicious add-on declares the same permissions as the victim app, thus, \emph{DiPrint} is not able to access to different resources than the expected ones. 
\subsection{Feature for Virtual Environment Detection: Package Name}

\textbf{Detection Mechanism 4.} Under VBA, the \emph{AntiPlugin App} is not installed on the device, which means that also the Android OS did not register it. Thus, the \emph{AntiPlugin App} queries the \texttt{PackageManager}~\cite{getInstalledApps} to get the list of all installed apps and check whether its package is in that list. Since \emph{Mascara} requires the victim app to be installed on the device, the \emph{AntiPlugin App} will receive a positive reply. 

\subsection{Feature for Virtual Environment Detection: App component name}

\textbf{Detection Mechanism 5.} The \texttt{getRecentTasks(int, int)}~\cite{getRecentTasks} and \texttt{getRunningTasks(int)}~\cite{getRunningTasks} APIs return the list of tasks recently launched by the user or currently running (a task is a collection of activities that users interact with when performing a certain job). Originally, any app could retrieve the list of tasks, even the ones referring to \texttt{Activities} of other apps. From Android level 21, those APIs were declared as deprecated and restricted to return only the caller's own tasks, i.e. only the list of \texttt{Activities} belonging to the app that called the API. Under VBA, the \emph{AntiPlugin App} can detect the tasks of other \emph{Plugin Apps} running in the same virtual environment. In \emph{Mascara}, we designed the \emph{Malicious APK} as an Android app declaring only \texttt{Services}, not \texttt{Activities}. Thus, even if calling the target API, the \emph{AntiPlugin App} will see only its own tasks. 

\textbf{Detection Mechanism 6.} The \texttt{getRunningServices(int)}~\cite{getRunningServices} API returns the list of running \texttt{Services}. Under VBA, the \emph{AntiPlugin App} might see \texttt{Services} belonging to other components running in the same virtual environment. By executing \emph{AntiPlugin} in \emph{Mascara}, we found out that this approach does not work properly.

\subsection{Feature for Virtual Environment Detection: Process Info}

\textbf{Detection Mechanism 7.} The \texttt{getRunningAppProcesses()}~\cite{getRunningAppProcesses} API returns the list of apps processes running on the device. If called by the \emph{AntiPlugin App} under a VBA, this API returns at least the \emph{Container App} process in addition to the \emph{AntiPlugin App}'s one. In \emph{Mascara}, all the processes under the same UID share the same name. However, through the \emph{Dynamic Proxy API Module} of DroidPlugin, we hooked the \texttt{getRunningAppProcesses()} method so that all process names are replaced with the victim app name.


\textbf{Detection Mechanism 8.} \emph{DiPrint} call the \texttt{Runtime.exec(String)} instruction with the \texttt{ps} argument to retrieve all processes information. In \emph{Mascara}, we used \emph{Whale} to hook the \texttt{Runtime.exec(String)} instruction and replace the \texttt{ps} argument with the \texttt{ls} one.  

\subsection{Feature for Virtual Environment Detection: Internal Storage Info}
For each newly installed app, the Android OS creates a private directory with the following path: \emph{/data/data/package\_name}. On the contrary, DroidPlugin assigns each \emph{Plugin App} a private directory with the following pattern: \emph{/data/data/container\_package\_name/Plugin/plugin\_package\_name}. 

\textbf{Detection Mechanism 9.} The \emph{AntiPlugin App} can check whether its private directory follows the default Android pattern. If not, the VBA can be detected. However, in \emph{Mascara} we hooked the \texttt{getApplicationInfo} API through the \emph{Dynamic Proxy API Module} of DroidPlugin to return the default Android directory pattern.

\textbf{Detection Mechanism 10.} Under VBA, the \emph{AntiPlugin App} can get the location of its APK by inspecting the \texttt{sourceDir} field of the \texttt{ApplicationInfo} object~\cite{applicationInfo}. Since the Android framework returns the \texttt{ApplicationInfo} for a given package name, in \emph{Mascara}, the \emph{AntiPlugin App} will retrieve the one referring to the victim app installed on the mobile device. 

\textbf{Detection Mechanism 11.} Under VBA, the process memory contains also the malicious APK path.
\emph{DiPrint} searches the existence of another different APK path in its own process memory by reading "\texttt{/proc/self/maps}". In \emph{Mascara}, any access to "\texttt{/proc/self/maps}" is disabled by hooking the constructor of the \texttt{File class} through \emph{Whale}.

\textbf{Detection Mechanism 12.} Android loads the dynamic-link library path from "\texttt{/data/app/package}". Under VBA dynamic-link libraries are loaded from a different path than the original. \emph{DiPrint} searches in the process memory the existence of sospicious paths. However, In \emph{Mascara}, any access to "\texttt{/proc/self/maps}" is disabled by hooking the constructor of the \texttt{File class} through \emph{Whale}.
\subsection{Feature for Virtual Environment Detection: Number of Launched App Activity and Service}

\textbf{Detection Mechanism 13.} By default, DroidPlugin declares only one stub \texttt{Service}. Under VBA, the \emph{AntiPlugin App} can launch a higher number of \texttt{Services} (or even other components) to trigger an exception. In \emph{Mascara}, the malicious add-on undergoes a customization process, through which all the victim app's components are copied into its \texttt{AndroidManifest.xml} file. Thus, the \emph{AntiPlugin App} has the exact number and type of expected components. 

\textbf{Detection Mechanism 14.} The \emph{AntiPlugin App} can inspect the number and type of components associated to its package name by analyzing the \texttt{PackageInfo} object~\cite{PackageInfo}. In \emph{Mascara}, the \emph{AntiPlugin App} will retrieve the  \texttt{PackageInfo} object associated to the victim app, already installed on the device. 

\subsection{Feature for Virtual Environment Detection: Static Broadcast Receiver}

\textbf{Detection Mechanism 15.} Among its stub components, DroidPlugin does not declare \texttt{Broadcast Receivers} since those are parsed from the \emph{AntiPlugin App}'s \texttt{AndroidManifest.xml} and dynamically registered at runtime. The \emph{AntiPlugin App} unregisters all the dynamically registered receivers and sends them an event to confirm that the process was successful. In \emph{Mascara}, all victim app components are copied int the malicious add-on, including \texttt{Broadcast Receivers}. 
\subsection{Feature for Virtual Environment Detection: Change App Component Property at Runtime}

\textbf{Detection Mechanism 16.} The \emph{AntiPlugin App} calls the \texttt{setComponentEnabledSetting(ComponentName, int, int)}~\cite{setComponentEnabledSetting} API to change the properties of a component at runtime. Under VBA, that call will trigger an error since the component will not be registered. By executing \emph{AntiPlugin} in \emph{Mascara}, we found out that this approach does not work properly.

\subsection{Feature for Virtual Environment Detection: Shared Native Components}

\textbf{Detection Mechanism 17.} Under VBA, the \emph{AntiPlugin App} might share the same virtual environment with other apps. If anyone has native components, such as \texttt{WebView}, these will share some internal data with other apps that have the same components. In \emph{Mascara}, we designed the malicious add-on so that the \emph{Malicious APK} does not have native components. 

\subsection{Feature for Virtual Environment Detection: Stack tracking of exception}

\textbf{Detection Mechanism 18.} \emph{DiPrint} analyzes the stack trace of the 13 lifecycle functions fir four components. Under VBA, the stack trace should be different. By executing \emph{DiPrint} in \emph{Mascara}, we found out that this approach does not work properly.

\section{Discussion}
\label{sec:discussion}
In this section, we illustrate the \emph{Mascara} limitations (Section~\ref{sec:limitations}) and the new detection mechanism we designed (Section~\ref{sec:countermeasures}). 

\subsection{Limitations}
\label{sec:limitations}
The first limitation consists in the execution of the victim app in a new \texttt{Context}, which is not its original one. When the malicious add-on loads the victim app classes and resources and executes them in the virtual environment, the victim app gains a new UID and looses the access to the data provided by the user until that time. Therefore, at the first execution, the victim app will loose its original state. The second limitation regards the startup time of the victim app, when launched for the first time inside the virtual environment, which might take few seconds (~6,64 sec on average for Telegram over ten attempts). Finally, the last limitation involves the creation of the shortcut in the Home screen, which could already exist on the user smartphone. 


\subsection{Countermeasures}
\label{sec:countermeasures}
\begin{figure}[h]
	\includegraphics[width=0.45\textwidth]{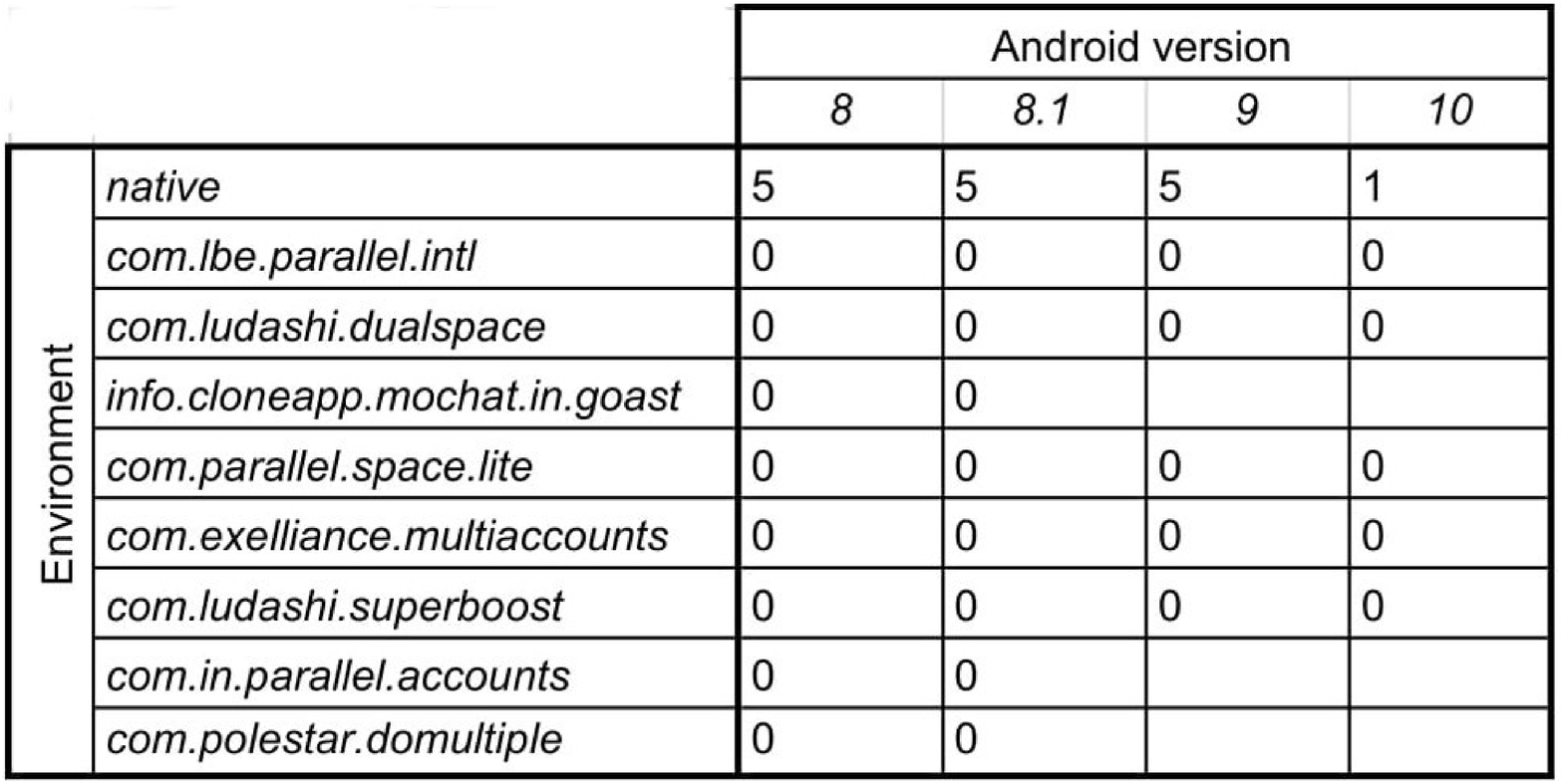}
	\caption{Comparison of \texttt{hotness\_count} values in Android native and virtual environments.}
	\label{fig:hotness_count_table}
\end{figure}

The existing defence mechanisms rely on the analysis of virtual environments properties retrieved through specific APIs, which could be easily modified by a malicious \emph{Container App}, that leverages on hooking techniques. This is the approach we used with \emph{Mascara}, thus, proving the ineffectiveness of existing defence mechanisms. On the contrary, to find a novel and robust countermeasure, we inspected the internal data structures of the Android Runtime (ART)~\cite{AndroidRuntime}, the Android apps execution environment. At the installation time, ART applies the Ahead-of-Time (AoT) compilation to compile Dalvik bytecode into binary code. Moreover, ART mirrors Java classes and methods of the Android framework and of the app custom code into C++ classes: the \texttt{Class} class and the \texttt{ArtMethod} class, respectively. At runtime, such classes are queried to retrieve the location of the binary code that needs to be executed. In Android 7.0, ART introduced a hybrid approach, based on both the Just-in-Time (JiT) compilation and the AoT one. Thus, the most frequently used methods are compiled AoT at installation time, while the remaining ones are interpreted. The key component of the hybrid compilation is the \texttt{hotness\_count} field of \texttt{ArtMethod} classes. This field provides the number of invocations and of loop iterations for a specific method, which is compiled according to the \texttt{hotness\_count} value. By analyzing the \texttt{hotness\_count} value for different methods in different Android versions, we found that a \texttt{hotness\_count} equal to 0 corresponds to the AoT. We, thus, chose the \texttt{hotness\_count} value of the \texttt{currentActivityThread} method in the \texttt{ActivityThread} class and evaluate its value when an app runs in the Android native environment and when it runs in a virtual environment. As shown in Fig.~\ref{fig:hotness_count_table}, the \texttt{hotness\_count} value is 0 for all the apps on the Play Store supporting the virtualization technique, which also means they all use the AoT compilation.  Based on this observation, we then developed a robust detection library (i.e., \emph{Singular}) that detects virtual environments by considering the \texttt{hotness\_count} value of \texttt{ArtMethod} classes. 

\section{Conclusions}
\label{sec:conclusion}
In this paper, we proposed \emph{Mascara}, a novel attack which relies on the virtualization technique to access and modify sensitive data available on the user's phone, as well as runtime data provided by the user. \emph{Mascara} involves a malicious add-on, which builds a virtual environment and runs both the victim app and a malicious APK inside it, gaining full control of the first one at runtime. We developed \emph{Mascarer}, a framework that automatically generates the malicious add-ons, and we tested it over 100 apps. We analyzed the runtime overhead introduced by the virtual environment and built up the \emph{Mascara} attack against Alamo, Amazon Music and Telegram. We, finally, identified a robust approach to determine whether an app is running in a virtual environment, by inspecting the \texttt{hotness\_count} value for \texttt{ArtMethod} classes in ART.

\balance

\bibliographystyle{IEEEtran}
\bibliography{mybiblio}

\newpage

\appendices
\onecolumn

\section{List of the Apps Used in the Experiments}
\label{appendix:A}

\begin{longtable}{|c|c|c|c|c|}
\hline
\textbf{\textbf{Index}} & \textbf{App Name}              & \textbf{Package Name}                             & \textbf{Size (MB)} & \textbf{N. Download} \\ \hline
\endfirsthead
\multicolumn{5}{c}%
{{\bfseries Table continued from previous page}} 

\\\hline
\textbf{\textbf{Index}} & \textbf{App Name}              & \textbf{Package Name}                             & \textbf{Size (MB)} & \textbf{N. Download}\\ \hline
\endhead
1              & 8 Ball Pool                    & com.miniclip.eightballpool                        & 13.86              & 500,000,000+        \\ \hline
2              & Alibaba.com                    & com.alibaba.intl.android.apps.poseidon            & 16.16              & 100,000,000+        \\ \hline
3              & Amazon Music                   & com.amazon.mp3                                    & 22.45              & 100,000,000+        \\ \hline
4              & Amazon Photos                  & com.amazon.clouddrive.photos                      & 9.62               & 10,000,000+         \\ \hline
5              & Amazon Prime Video             & com.amazon.avod.thirdpartyclient                  & 39.56              & 100,000,000+        \\ \hline
6              & Amazon Shopping                & com.amazon.mShop.android.shopping                 & 2.97               & 100,000,000+        \\ \hline
7              & Aquapark.io                    & com.cassette.aquapark                             & 11.68              & 100,000,000+        \\ \hline
8              & Art of War: Legions            & com.addictive.strategy.army                       & 29.43              & 10,000,000+         \\ \hline
9              & Audible Audiobooks             & com.audible.application                           & 31.51              & 100,000,000+        \\ \hline
10             & Ball Pass 3D                   & com.Kjoeb.BallPass                                & 13.3               & 1,000,000+          \\ \hline
11             & BIGO Live                      & sg.bigo.live                                      & 31.6               & 100,000,000+        \\ \hline
12             & Border patrol                  & com.fivebits.borderpatrol                         & 11.71              & 10,000,000+         \\ \hline
13             & Brain Out                      & com.mind.quiz.brain.out                           & 10.07              & 100,000,000+        \\ \hline
14             & Brain Test                     & com.unicostudio.braintest                         & 4.77               & 50,000,000+         \\ \hline
15             & Burger King APP                & com.emn8.mobilem8.nativeapp.bk                    & 9.5                & 1,000,000+          \\ \hline
16             & Calm – Meditate,Sleep,Relax    & com.calm.android                                  & 26.81              & 10,000,000+         \\ \hline
17             & Candy Crush Saga               & com.king.candycrushsaga                           & 7.35               & 1.000.000.000+      \\ \hline
18             & Candy Crush Soda Saga          & com.king.candycrushsodasaga                       & 7.62               & 100,000,000+        \\ \hline
19             & Chick-fil-A                    & com.chickfila.cfaflagship                         & 15.87              & 10,000,000+         \\ \hline
20             & Cisco Webex Meetings           & com.cisco.webex.meetings                          & 16.23              & 10.000.000+         \\ \hline
21             & Coin Master                    & com.moonactive.coinmaster                         & 11.44              & 100,000,000+        \\ \hline
22             & Cut and Paint                  & com.painter.cutandpaint                           & 18.9               & 10,000,000+         \\ \hline
23             & Dancin Road                    & com.amanotes.pamadancingroad                      & 14.49              & 50,000,000+         \\ \hline
24             & Dentist Bling                  & com.crazylabs.dentist                             & 24.26              & 10,000,000+         \\ \hline
25             & Dig This!                      & se.raketspel.digaround                            & 24.22              & 10.000.000+         \\ \hline
26             & Discord                        & com.discord                                       & 23.01              & 100,000,000+        \\ \hline
27             & Draw Climber                   & com.appadvisory.drawclimber                       & 23.04              & 50.000.000+         \\ \hline
28             & Emoji Home                     & com.home.emoticon.emoji                           & 27.82              & 1,000,000+          \\ \hline
29             & Enlight Pixaloop               & com.lightricks.pixaloop                           & 12.8               & 10.000.000+         \\ \hline
30             & Epic Race 3D                   & com.gym.racegame                                  & 18.91              & 10.000.000+         \\ \hline
31             & Extreme Car Driving Simulator  & com.aim.racing                                    & 11.94              & 100,000,000+        \\ \hline
32             & Funimation                     & com.Funimation.FunimationNow.androidtv            & 17.44              & 50,000+             \\ \hline
33             & Gardenscapes                   & com.playrix.gardenscapes                          & 11.92              & 100.000.000+        \\ \hline
34             & GrubHub                        & com.grubhub.android                               & 21.18              & 10,000,000+         \\ \hline
35             & HBO Max                        & com.hbo.hbonow                                    & 8.11               & 10.000.000+         \\ \hline
36             & Hide ‘N Seek 3D                & helperapp.helper.hidenseekhelper                  & 7.15               & 5,000,000+          \\ \hline
37             & Hitmasters                     & com.playgendary.hitmasters                        & 20.23              & 10,000,000+         \\ \hline
38             & Hole.io                        & io.voodoo.holeio                                  & 14.62              & 50,000,000+         \\ \hline
39             & Home Restoration               & com.panteon.homesweethome                         & 14.42              & 5.000.000+          \\ \hline
40             & Homecapes                      & com.playrix.homescapes                            & 9.74               & 100.000.000+        \\ \hline
41             & Houseparty                     & com.herzick.houseparty                            & 23.19              & 10.000.000+         \\ \hline
42             & Hulahoop                       & com.hulahoop.android                              & 18.17              & 10,000,000+         \\ \hline
43             & Hungry Shark Evolution         & com.fgol.HungrySharkEvolution                     & 5.21               & 100.000.000+        \\ \hline
44             & I, The One – Action Fighting   & vh.one                                            & 26.93              & 10,000,000+         \\ \hline
45             & ID Please – Club Simulation    & com.NeverEndingGames.IdPlease                     & 25.51              & 10,000,000+         \\ \hline
46             & iHeartRadio                    & com.clearchannel.iheartradio.controller           & 47.05              & 50,000,000+         \\ \hline
47             & Johnny Trigger                 & com.time.trigger                                  & 14.82              & 50.000.000+         \\ \hline
48             & KeepClean lite                 & com.appsinnova.android.keepclean.lite             & 16.52              & 100,000+            \\ \hline
49             & Klondike Adventures            & com.vizorapps.klondike                            & 22.61              & 10,000,000+         \\ \hline
50             & Likee lite                     & video.like.lite                                   & 15.43              & 50,000,000+         \\ \hline
51             & Little Caesars                 & com.littlecaesars                                 & 10.27              & 5.000.000+          \\ \hline
52             & Micosoft Teams                 & com.microsoft.teams                               & 21.98              & 50,000,000+         \\ \hline
53             & Mr Ninja                       & com.lionstudios.mrninja                           & 24.51              & 10.000.000+         \\ \hline
54             & MyBoost                        & com.boost.care                                    & 15.62              & 1.000.000+          \\ \hline
55             & NERF Epic Pranks!              & games.nerf.epic.pranks.free                       & 3.36               & 10,000,000+         \\ \hline
56             & News Break                     & com.particlenews.newsbreak                        & 11.75              & 10,000,000+         \\ \hline
57             & News Home                      & com.home.news.breaking                            & 38.19              & 1,000,000+          \\ \hline
58             & Norton Security VPN            & com.symantec.securewifi                           & 16.22              & 10,000,000+         \\ \hline
59             & One Booster                    & com.cleanteam.oneboost                            & 17.5               & 10,000,000+         \\ \hline
60             & Onnect - Pair Matching Puzzle  & com.gamebility.onet                               & 16.8               & 10,000,000+         \\ \hline
61             & Padenatef                      & com.sun.newjbq.beijing.ten                        & 28.26              & 500,000+            \\ \hline
62             & Paint By Number                & paint.by.number.pixel.art.coloring.drawing.puzzle & 15.16              & 50,000,000+         \\ \hline
63             & PictureThis                    & cn.danatech.xingseus                              & 26.36              & 5.000.000+          \\ \hline
64             & Pinterest Lite                 & com.pinterest.twa                                 & 2.97               & 1,000,000+          \\ \hline
65             & Pizzaiolo!                     & com.pizza.dough                                   & 17.96              & 1,000,000+          \\ \hline
66             & Postmates Merchant App         & com.postmates.android.merchant                    & 24.38              & 100,000+            \\ \hline
67             & Recipes Home                   & com.home.recipes.food.drink                       & 27.94              & 1,000,000+          \\ \hline
68             & Rescue Cut                     & com.app.rescuecut                                 & 13.57              & 50,000,000+         \\ \hline
69             & Roblox                         & com.roblox.client                                 & 24.33              & 100,000,000+        \\ \hline
70             & Roku Remote Control            & com.tinybyteapps.robyte                           & 7.37               & 5,000,000+          \\ \hline
71             & Shein                          & com.zzkko                                         & 33.96              & 100,000,000+        \\ \hline
72             & SiriusXM                       & com.sirius                                        & 10.52              & 10,000,000+         \\ \hline
73             & Skype                          & com.skype.raider                                  & 9.29               & 1.000.000.000+      \\ \hline
74             & Slap Kings                     & mobi.gameguru.slapkings                           & 11.56              & 50,000,000+         \\ \hline
75             & SmartNews                      & jp.gocro.smartnews.android                        & 18.39              & 10.000.000+         \\ \hline
76             & Soap Cutting                   & com.crazylabs.soap.cutting                        & 23.84              & 10,000,000+         \\ \hline
77             & SONIC Drive-In                 & com.sonic.sonicdrivein                            & 72.37              & 5,000,000+          \\ \hline
78             & Sort it 3D                     & com.game.sortit3d                                 & 15.05              & 10.000.000+         \\ \hline
79             & Spotify                        & com.spotify.music                                 & 26.48              & 500,000,000+        \\ \hline
80             & Super Salon                    & com.flashreadinc.pluckinawesome                   & 22.95              & 5,000,000+          \\ \hline
81             & TBN: Watch TV                  & tbn\_mobile.android                               & 23.54              & 500,000+            \\ \hline
82             & The Chosen                     & com.vidangel.thechosen                            & 22.96              & 1,000,000+          \\ \hline
83             & The Cook                       & com.pd.thecook                                    & 14.76              & 10,000,000+         \\ \hline
84             & Tiles Hop                      & com.amanotes.beathopper                           & 23.57              & 100,000,000+        \\ \hline
85             & Tubi TV                        & com.tubitv                                        & 19.37              & 50.000.000+         \\ \hline
86             & Ultimate Disc                  & com.malvo.frisbee                                 & 16.46              & 10.000.000+         \\ \hline
87             & US breaking news               & ma.safe.bnus                                      & 13.9               & 1.000.000+          \\ \hline
88             & Video Editor \& Video Maker    & com.camerasideas.instashot                        & 13.33              & 100.000.000+        \\ \hline
89             & Walmart                        & com.walmart.android                               & 17.58              & 50,000,000+         \\ \hline
90             & Walmart Grocery                & com.walmart.grocery                               & 30.22              & 50.000.000+         \\ \hline
91             & Water Race 3D                  & com.ihd.waterrace                                 & 15.86              & 1,000,000+          \\ \hline
92             & Webtoon                        & com.naver.linewebtoon                             & 13.16              & 50,000,000+         \\ \hline
93             & Whatsapp Messenger             & com.whatsapp                                      & 11.56              & 5,000,000,000+      \\ \hline
94             & Wish                           & com.contextlogic.wish                             & 7.2                & 100,000,000+        \\ \hline
95             & Wood Block Puzzle              & com.fastfun.tetris                                & 17.48              & 1,000,000+          \\ \hline
96             & Woodturning                    & com.BallGames.Woodturning                         & 23.94              & 50,000,000+         \\ \hline
97             & Word Crush                     & com.tangramgames.gourddoll.wordcrush              & 13.74              & 1.000.000+          \\ \hline
98             & Wordscapes                     & com.peoplefun.wordcross                           & 17.22              & 10,000,000+         \\ \hline
99             & Yahoo Mail                     & com.yahoo.mobile.client.android.mail              & 26.51              & 100,000,000+        \\ \hline
100            & Your Phone Companion           & com.microsoft.appmanager                          & 24.96              & 100.000.000+        \\ \hline
\end{longtable}

\end{document}